\documentclass[3p]{elsarticle}

\usepackage{hyperref}
\usepackage[boxruled,vlined]{algorithm2e}
\usepackage{amssymb,amsmath}
\usepackage{lipsum}
\usepackage{multirow}
\usepackage{url,enumitem,bm}
\usepackage{xcolor}
\journal{Applied Acoustics}

\DeclareTextCommandDefault{\textcopyright}{\textcircled{c}}
\DeclareTextCommandDefault{\textregistered}{\textcircled{%
      \check@mathfonts\fontsize\sf@size\z@\math@fontsfalse\selectfont R}}

\begin{document}

{\color{blue}This manuscript has been accepted for publication at Applied Acoustics. Once published, this arcticle can be found at \url{www.journals.elsevier.com/applied-acoustics}.}\\

\vspace{15cm}

{\color{blue}\textcopyright 2019. This manuscript version is made available under the CC-BY-NC-ND 4.0 license \url{http://creativecommons.org/licenses/by-nc-nd/4.0/}}

\newpage
\begin{frontmatter}

\title{Directional Embedding Based Semi-supervised Framework For Bird Vocalization Segmentation}

\author[add1]{Anshul Thakur}
\ead{anshul\_thakur@students.iitmandi.ac.in}
\author[add1]{Padmanabhan Rajan}
\ead{padman@iitmandi.ac.in}
\address[add1]{School of Computing and Electrical Engineering, \\Indian Institute of Technology Mandi,\\ Himachal Pradesh, India}

\begin{abstract}
This paper proposes a data-efficient, semi-supervised, two-pass framework for segmenting bird vocalizations. The framework utilizes a binary classification model to categorize frames of an input audio recording into the background or bird vocalization. The first pass of the framework automatically generates training labels from the input recording itself, while model training and classification is done during the second pass. The proposed framework utilizes a reference directional model for obtaining a feature representation called directional embeddings (DE). This reference directional model acts as an acoustic model for bird vocalizations and is obtained using the mixtures of Von-Mises Fisher distribution (moVMF). The proposed DE space only contains information about bird vocalizations, while no information about the background disturbances is reflected. The framework employs supervised information only for obtaining the reference directional model and avoids the background modeling. Hence, it can be regarded as semi-supervised in nature. The proposed framework is tested on approximately 79000 vocalizations of seven different bird species. The performance of the framework is also analyzed in the presence of noise at different SNRs. Experimental results convey that the proposed framework performs better than the existing bird vocalization segmentation methods.
\end{abstract}

\begin{keyword}
Bird vocalization segmentation\sep Bioacoustics\sep Directional Embedding
\end{keyword}

\end{frontmatter}

\section{Introduction}
\label{sec:1}
Birds are of significant importance in maintaining the equilibrium of various ecosystems. However, many bird species are facing the threat of population decline due to habitat destruction and global climate change \cite{brandes2008automated}. As a result, there is a pressing need to study the effects of human activities and climate change on avian diversity. Surveying and monitoring of bird species in their natural habitat are generally the first steps in analyzing these effects \cite{lee2008}. Many bird species are highly vocal, hence, acoustic monitoring provides a passive and convenient way to monitor birds. The development of sophisticated programmable recording devices has made acoustic monitoring feasible. Acoustic data recorded by these devices can be analyzed to perform various tasks such as species identification, the tracking of migrant species or in examining the avian biodiversity of a given region. Manual processing of this data is tedious and time-consuming, and requires considerable knowledge of bird vocalizations. Hence, it is essential to develop automatic methods for processing the recorded acoustic data.   

Segmenting bird vocalizations and discarding the background is the principal step in many bioacoustic tasks such as studying the semantics of bird songs or species identification. The performance of this segmentation step directly influences the overall performance of other subsequent tasks \cite{fagerlund_thesis}. Bird vocalization segmentation appears to be a specialized version of sound event detection (SED). These two tasks differ from each other in nature of the target acoustic events. The typical target events, such as gun-shot, baby-cry, glass-break and dog-bark, in any SED task (such as the DCASE challenges \cite{dcase}) are well-defined i.e. the knowledge about the temporal-frequency structure of the target events is present as prior information. However, in bird vocalization segmentation, the nature of the target event (bird vocalizations) is not well defined. This can be attributed to the large amount of variations present in the frequency profiles of vocalizations of different species. This task of bird vocalization segmentation is often challenging in the field conditions due to reasons listed below:

\begin{itemize}
    \item \emph{\textbf{Background modeling}}: The presence of various biotic (e.g. vocalizations of other animals) and abiotic (such as rain, wind or human-induced sounds) background sounds make it difficult to isolate the bird vocalizations from the background. Irrespective of the source, these background disturbances are unpredictable and highly volatile. This makes the acoustic modeling of the background non-trivial.
    
    \item \emph{\textbf{Bird vocalization modeling}}: The vocalizations of different bird species can be significantly different from each other. For example, the temporal-frequency modulations present in Cassin's vireo (a North American song bird) vocalizations are completely different from the temporal-frequency modulations present in vocalizations of Greater sooty owl and Hooded crow. Apart from that, many species exhibit a large repertoire of vocalization types, which results in high intra-class variations. Hence, coming up with one universal acoustic model that can be used for segmenting the vocalizations of many bird species is not straight-forward. 
     
\item \emph{\textbf{Lack of labeled training data}}: Like other bioacoustic tasks, bird vocalization segmentation also suffers from the scarcity of labeled training data. This limits the use of data-intensive state-of-art audio classification frameworks such as convolutional neural networks (CNN).  In recent times, large datasets with weak labels (i.e. presence or absence of birdcalls) have been released for the audio tagging tasks (also known as \emph{bird activity detection}) \cite{bad}. These datasets lack vocalization-level time-stamps or frame-level labels and can be used for training powerful neural networks in a weakly supervised setup \cite{large_scale} for segmenting bird vocalizations. However, the segmentation performance of these weakly supervised neural networks is not on-par with their supervised counterparts.     
  
\end{itemize}

In this work, we propose a data-efficient semi-supervised two-pass framework which overcomes the aforementioned challenges associated with the task of bird vocalization segmentation. The proposed segmentation framework does not require any explicit background acoustic model and utilizes a reference bird vocalization acoustic model to discriminate between bird and non-bird sounds. This reference vocalization model is obtained using a very small amount of the labeled training data. Since only bird vocalizations are modeled (not the background), the proposed framework can be considered as a semi-supervised framework. This notion of semi-supervision is borrowed from non-negative matrix factorization (NMF) based speech enhancement literature where either speech samples or noise is used for training \citep{nmf,mld}.

\begin{figure}[t]
	\centering
	\includegraphics[trim={20cm 1cm 19cm 0cm},scale=0.75]{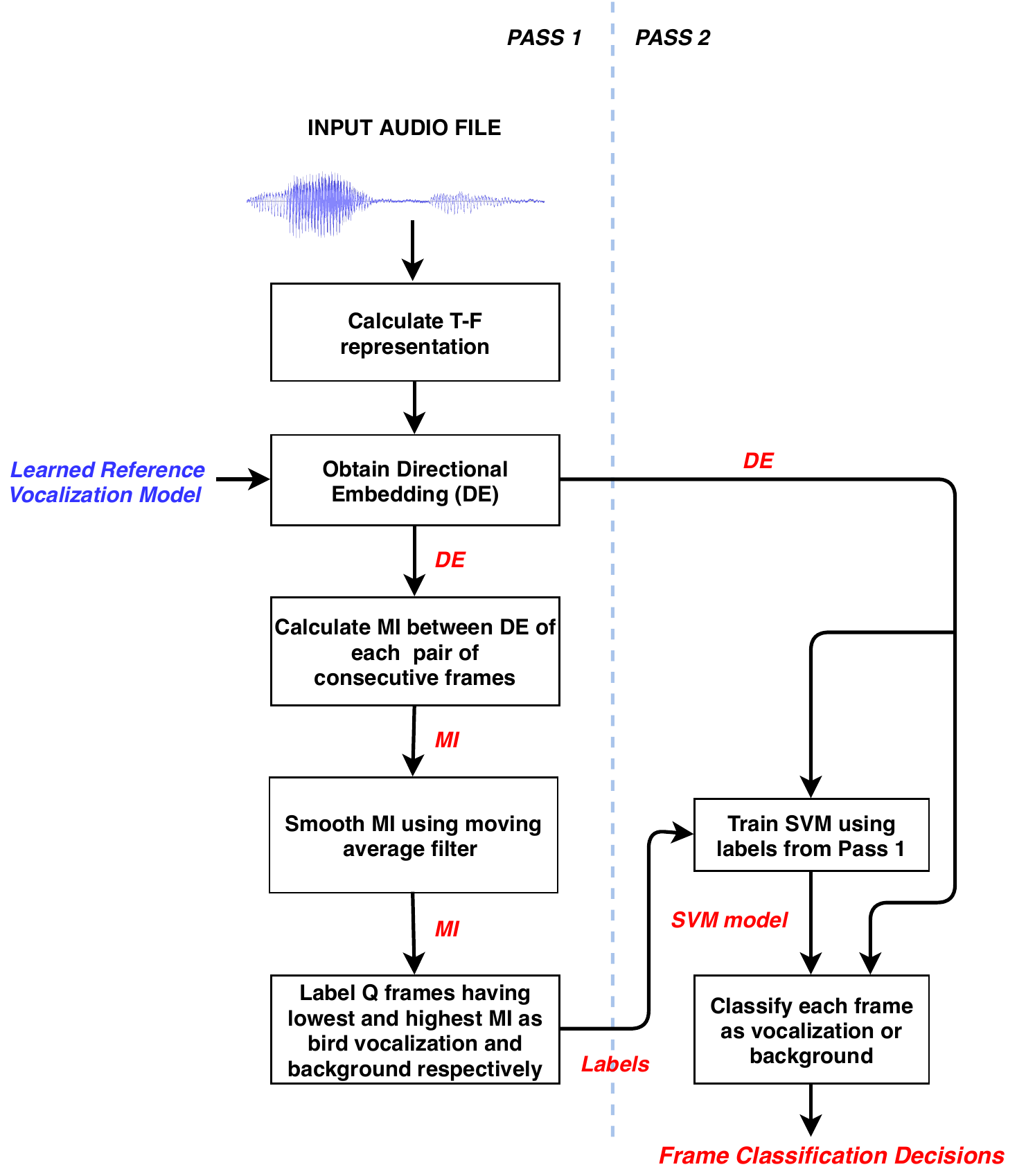}
	\caption{Proposed semi-supervised framework for bird vocalization segmentation}
	\label{fig:framework}
\end{figure}

Given a reference model for bird vocalizations, the proposed framework is a two-pass process. During first pass, the reference vocalization model is utilized to convert the input audio recording into a frame-wise feature representation called directional embeddings (DE). The feature space corresponding to DE retains information only about the bird vocalizations and most of the background disturbances are inherently not reflected in the feature space (more details are in Section \ref{sec:features_learning}). Mutual information (MI) is obtained between consecutive pairs of DE frames to get an initial estimate of the regions containing the background and vocalizations respectively. MI exhibits high values for the background and low values for vocalization regions (see Section \ref{sec:proposed}). A fixed number of audio frames exhibiting highest and lowest MI are then automatically labeled as background and vocalizations respectively. These labels and DE features are given as input to the second pass of the framework. In the second pass, a classification model is trained on the DE features using the generated labels. This trained classification model is utilized to classify each input audio frame as background or bird vocalization. Fig.~\ref{fig:framework} illustrates the proposed two-pass semi-supervised framework for bird vocalization segmentation, where a support vector machine (SVM) is used as the classification model in the second pass.

DE forms the nucleus of the proposed framework. As discussed earlier, DE provides ability to discriminate bird sounds from non-bird background disturbances. Hence, the utilization of DE in the proposed framework helps in decreasing false alarm rate significantly. DE are obtained by projecting the time-frequency representations of an input audio on a dictionary (which forms the reference vocalization model).
The atoms of this dictionary are unit vectors pointing in the dominant directions of bird vocalizations. These dominant directions are learned by clustering a small amount of audio frames (henceforth termed as training data) containing bird vocalizations on the unit hypersphere. Clustering is achieved using mixtures of Von-Mises Fisher distribution (moVMF) \citep{banerjee} and the mean directional vector of each cluster forms an atom of the dictionary. Also, moVMF can be seen as equivalent of Gaussian mixture models (GMM) on the unit hypersphere. Clustering the spectral frames on the unit hypersphere helps in capturing the information about the presence/absence of dominant frequency bins rather than the frequency magnitudes. This mitigates the effect of near-field and far-field conditions on the clustering process to some extent. Hence, two similar vocalizations of a bird species, recorded at different distances from the recording device, are clustered together. This behaviour is not present in other clustering techniques such as K-means where magnitude information influences the assignment of a data point to any cluster. Instead of moVMF, earlier studies \cite{dan_skm,multi_skm} have used spherical K-means to cluster the data for audio feature learning.
However, moVMF has been reported to provide better clustering than spherical K-means \cite{zhong}. The other advantage of moVMF is the presence of the concentration parameter (see section \ref{sec:features_learning}) which helps in discarding mean vectors of clusters having very low concentration around the mean. This helps in overcoming the effects of outliers which can be introduced in the learning process of moVMF due to incorrect labels in the training data.

The main contributions of this work are listed below:

\begin{itemize}
    \item A semi-supervised two pass framework which can be used for segmenting vocalizations of many bird species in various recording environments. The framework does not require any background modeling and utilizes a very small amount of training data to learn a bird vocalization acoustic model. Also, the framework avoids the train-test conditions mismatch by using a classifier trained on the DE features extracted from the input recording itself.   
    
    \item Treating the audio frames as directional data to mitigate the effect of near-field and far-field conditions. This directional data is obtained by projecting the audio frames on the unit hypersphere.
    
    \item Utilizing moVMF as an acoustic model to capture the behaviour of a particular class (in this case, bird vocalizations). To the best of authors' knowledge, this is the first study to treat moVMF as an acoustic model.  
    
    \item An intermediate feature representation (directional embeddings) suitable to facilitate bird vocalization segmentation in a semi-supervised manner. This feature representation equips the proposed framework with the ability to discriminate between bird and non-bird sounds.  
\end{itemize}

The rest of the paper is organized as follows: In section \ref{sec:background}, various existing studies targeting bird vocalization segmentation and acoustic event segmentation are discussed. The proposed directional embeddings based feature learning is described in section \ref{sec:features_learning}. The proposed framework is discussed in section \ref{sec:proposed}. Later, we discuss experimental details, results and conclusion in sections \ref{sec:exp1}, \ref{sec:results} and \ref{Sec:con} respectively.

\section{Previous studies}
\label{sec:background}
The literature contains many studies addressing the task of bird vocalization segmentation. These studies can be categorized into the following:

\textbf{\textit{Manual segmentation:}} A few early bioacoustic studies dealing with bioacoustics \cite{Trifa, lee2008} have used manually segmented bird vocalizations. Manual segmentation can be tedious and unfeasible if large amount of audio data has to be processed.

\textbf{\textit{Segmentation using time-domain features:}} Segmentation methods based on time-domain features such as energy, entropy and inverse spectral flatness (ISF) have been used for segmenting bird vocalizations. Many studies\cite{Harma,Somervuo,Fagerlund} threshold the energy envelope to extract syllables from bird song recordings. Spectral entropy \cite{wang2013} and KL-divergence \cite{Lakshmi} have also been used to extract bird vocalizations from audio recordings. The regions containing bird vocalizations tend to exhibit lower entropy in comparison to the background regions. Also, KL-divergence between normalized power spectral density (PSD) of an audio frame and the uniform density is an indicator of entropy; lower KL-divergence corresponds to high entropy and vice-versa. Lakshminarayanan \emph{et. al} \cite{Lakshmi} exploited this behavior for segmenting bird vocalizations. In one of our earlier studies, we used short term energy (STE) and inverse spectral flatness (ISF)  for segmentation \cite{thakur2016}.

All these methods are unsupervised in nature which is desirable. However, these methods are not able to discriminate bird vocalizations from any other sound event. 
Also, the performance of energy/ISF based methods rapidly deteriorate as the SNR decreases.

\textbf{\textit{Methods based on spectrogram processing:}}
Various spectrogram processing based methods \cite{Oliveira,fodor,las,glotin} have been proposed in the literature for bird vocalization segmentation. These methods have a common theme. First, morphological processing is applied on spectrograms to remove the low-energy background acoustic events. Then, thresholding is applied on the sum of frequency magnitudes of each frame of the morphologically processed spectrogram to segment the vocalizations. Apart from these methods, Koluguri \emph{et al.} \cite{koluguri} proposed to use directional filtering on a spectrogram enhanced using a multiple window Savitzky-Golay filter. This enhancement procedure can enhance far-field sounds which are often ignored by many other methods. The directional filtering is applied in four different directions to capture the time-frequency modulations of bird vocalizations. Like time-domain based methods, these spectrogram based methods are also not able to distinguish bird vocalizations from other background sounds and output any high-energy acoustic event as bird vocalization.   


\textbf{\textit{Other Methods:}}
A noise-robust template matching based method is proposed by Kaewtip \emph{et.~al} \cite{templates}. This method performs well in challenging conditions. However, the major disadvantage of template matching is that we must know beforehand what vocalizations we wish to segment. Hence, this method may not scale in real-world scenarios. Neal \emph{et.~al} \cite{Neal} proposed a noise robust bird vocalization segmentation method which uses a random forest classifier to yield a probability mask from the spectrograms. A hard threshold is applied on this probability mask to generate the binary decision i.e. bird/non-bird at pixel level of the spectrogram. The main disadvantage of this method is that it requires large amount of labeled training data which is usually scarce.

In our earlier studies \cite{thakur2016, thakur_ncc}, we proposed model-based unsupervised frameworks for segmentation. These frameworks bypasses the need of thresholding by using a classification model. However, like other existing methods, these frameworks are also not able to distinguish between bird and non-bird sounds. We also proposed a semi-supervised framework \citep{thakur_mlsp} that utilizes a singular value decomposition (SVD) based feature representation to discriminate between bird/non-bird sounds. The ideal number of singular vectors required to obtain maximum discrimination between bird and non-bird sounds depend on the species under study. This limits the effectiveness of this method in field conditions.

\section{Feature Learning : Directional Embedding}
\label{sec:features_learning}
Feature learning, in the context of bird vocalization segmentation, deals with finding a representation, which not only highlights the bird vocalizations but also removes background sounds. In this section, we describe the proposed feature representations called directional embedding (DE) for segmenting bird vocalizations. Von Mises-Fisher mixture model (moVMF), which is based on directional statistics, is used to learn a dictionary whose atoms correspond to the dominant directions of bird vocalizations. The mean directions of mixtures of vMF are regarded as dominant directions. As discusses in Section \ref{sec:1}, the DE representation is obtained by projecting the spectrogram of an audio recording on this dictionary. Here, first we explain the pre-processing step to convert the input audio recordings into a super-frame representation. Then, we describe moVMF to obtain the dictionary of dominant directions (the reference vocalization model). Later, the method to obtain DE from an audio recording using the trained moVMF is explained. A discussion comparing DE with features obtained from NMF is also included in this section. The overall process to obtain DE is documented in Algorithm \ref{alg:algo1}.

\begin{algorithm}
\label{alg:algo1}
\caption{Procedure to obtain the proposed DE representation}  
\textbf{\textit{Training: Obtaining dominant directions of bird vocalizations}}\\
\textbf{Input:} Training audio recordings. \textbf{Output:} A dictionary of the mean directions.
	\begin{itemize}[rightmargin=\dimexpr\linewidth-10.5cm-\leftmargin\relax]
	\setlength\itemsep{0.05em}
    \item Obtain magnitude spectrograms of all the training audio recordings.
    \item Convert each spectrogram into the super-frame based representation
    and project each super-frame on the unit hypersphere as explained in Section \ref{sec:features_learning}.
    \item Use the available ground truth of the training audio recordings to extract and pool the unit norm super-frames corresponding to the bird vocalization regions. The background super-frames are simply discarded. 
    
    \item Learn a moVMF having $Z$ mixtures on the selected unit norm super-frames.  
    
    \item Form a dictionary, $\mathbf{M}$ having $Z$ atoms, whose atoms are the mean directions of the mixtures of VMF (moVMF). 
    \end{itemize}
\textbf{\textit{Feature learning: Obtaining DE for any audio recording}} 

\textbf{Input:} An audio recording and the dictionary of the mean directions. \textbf{Output:} DE features.
\begin{itemize}
    \item Calculate the spectrogram from any input audio recording.
    
    \item Process this spectrogram to obtain super-frames.
    
    \item Project these super-frames on the dictionary $\mathbf{M}$ (learned during training) to obtain the proposed DE representation.
  
\end{itemize}

\end{algorithm}

\subsection{Pre-processing audio recordings}
\label{ssec:prep}
Most bird vocalizations exhibit temporal and frequency (T-F) modulations \cite{marler}. These modulations give specific character to a bird vocalization. However, when the short-term Fourier transform is applied to obtain the spectrogram of an audio recording, the information about T-F modulations is smeared out as one frame is not capable of capturing these modulations effectively. These T-F modulations are not only helpful in classifying bird sounds but also provides important cues to discriminate them from other background sounds. Hence, we embed the temporal context around each frame to capture these modulations. A moving window of $w$ frames with a stride of a single frame is used to embed this temporal context. The $w$ frames within the moving window are concatenated, one below the other, to form a super-frame. If the T-F representation has dimensions of $d \times n$ ($d$ frequency bins and $n$ frames), then after applying concatenation process, a $wd \times n$ dimensional super-frame based representation is obtained. This representation is used for both learning the dictionary and for segmenting the input recordings. 
\begin{figure}[t]
	\centering
	\includegraphics[width=0.75\textwidth,height= 5.5 cm]{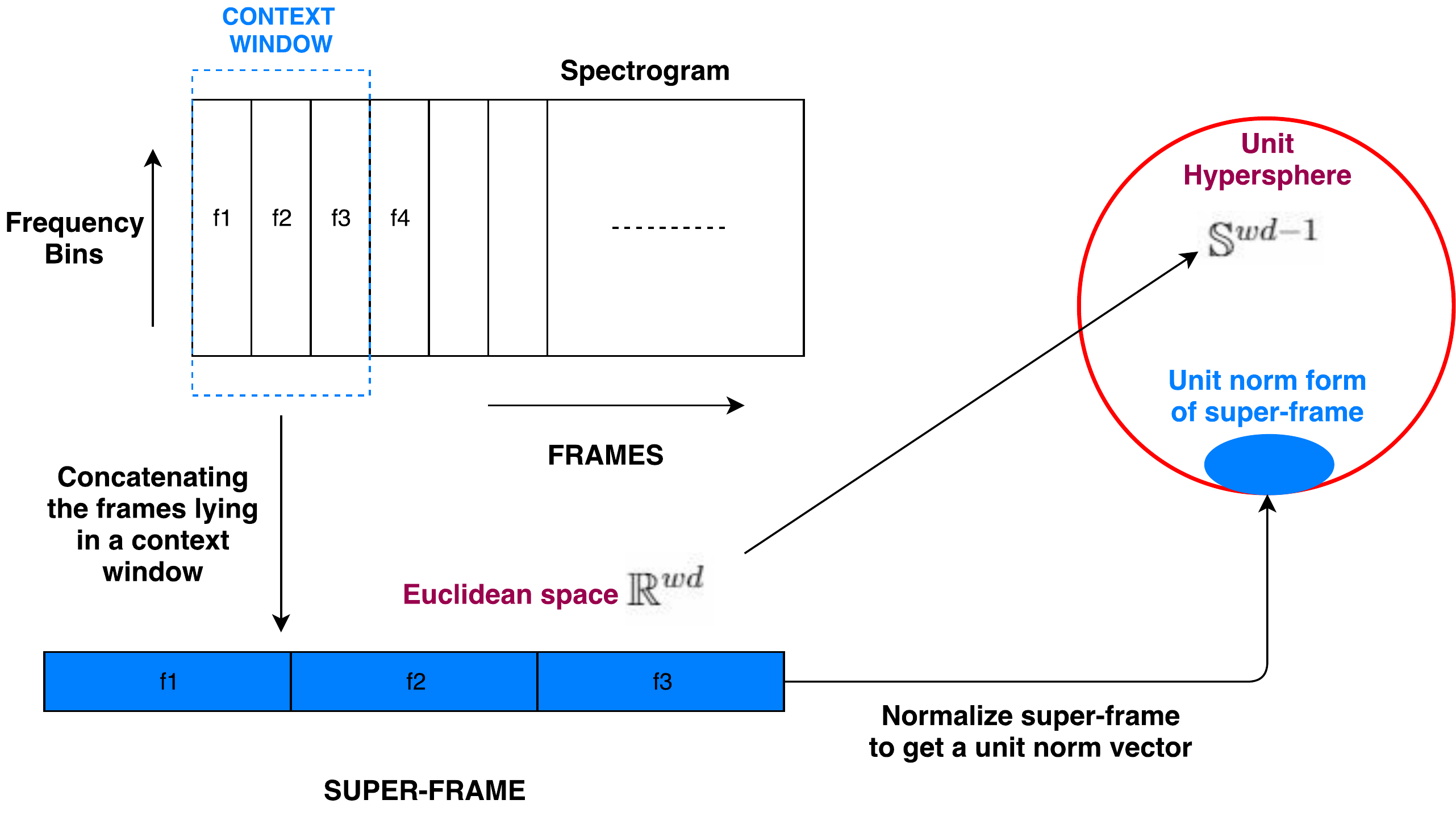}
	\caption{Schematic representation of the process to convert a spectrogram into the unit norm based representation}
	\label{fig:preprocessing}
\end{figure}

For learning the dictionary, we are interested in clustering the unit norm representations of super-frames on the unit hypersphere. Therefore, we normalize each training super-frame to get a unit norm vector that lies on the surface of the unit hypersphere. The magnitude information is lost due to this transformation. However, the desired information about dominant frequencies is still preserved. Fig.~\ref{fig:preprocessing} illustrates the process of converting any input spectrogram into a unit norm based representation.

\subsection{Mixtures of von Mises-Fisher distribution (moVMF)}
The super-frames obtained by concatenation of $w$ $d$-dimensional frames lies in the Euclidean space, $\mathbb{R}^{wd}$ and the unit norm variants of these super-frames lie on surface of the unit hypersphere, $\mathbb{S}^{wd-1}$. Mixtures of von Mises-Fisher distribution \cite{banerjee} can be used as clusters of these unit norm vectors. moVMF has found its application across many fields in various applications such as speaker clustering \citep{tang2009}, medical imaging \citep{bangert}, text mining \citep{text} and estimation of shadow and illumination in computer vision \citep{shadow_cvpr}. Taking a cue from these studies, we propose to use vMF mixture modeling to cluster unit norm representations of the super-frames. Originally two variants of moVMF were proposed: soft assignments and hard assignments \citep{banerjee}. In this work, we utilize moVMF with soft assignments only.

To learn moVMF as the bird vocalization model, the training audio recordings are manually segmented, using the available ground truth, to extract the bird vocalization regions. The super-frames corresponding to these bird vocalization regions are pooled together. Suppose $\mathcal{X}=\{\mathbf{x}_1,\mathbf{x}_2,.....,\mathbf{x}_N\}$ is a set of unit norm representations of these pooled super-frames where $\mathbf{x}_i \in \mathbb{R}^{wd}$, $\Vert \mathbf{x}_i \Vert=1$ and thus $ \mathbf{x}_i \in \mathbb{S}^{wd-1}$. The $\mathbf{x}_i$ has $wd$-variate vMF distribution if its probability density is \cite{banerjee,gopal}: 

 \begin{equation}
 \label{eq:vmf}
\rho(\mathbf{x}_i;\bm{\mu},\kappa)=C_{wd}(\kappa)\exp^{\kappa\bm{\mu}^T\mathbf{x}_i},   
 \end{equation}

where $\Vert\bm{\mu}\Vert=1$, $\kappa>0$ and $wd\geq2$. $C_{wd}(\kappa)$ is a normalizing constant defined as:
 \begin{equation}
 \label{eq:nmc}
   C_{wd}(\kappa)=\frac{\kappa^{(wd/2)-1}}{(2\pi)^{wd/2}I _{wd/2-1}(\kappa)}.
 \end{equation}

Here $I _{wd/2-1}(.)$ is a modified Bessel's function of first kind and of order $wd/2-1$. The density function, $\rho(\mathbf{x}_i;\bm{\mu},\kappa)$, is parameterized by mean direction, $\boldsymbol{\mu}$, and concentration parameter, $\kappa$ which characterizes the concentration of unit norm vectors around the mean direction. As $\kappa \to 0$, $\rho(\mathbf{x}_i;\bm{\mu},\kappa)$ becomes the uniform density on the surface of the unit hypersphere. On the other hand, when $\kappa \to \infty$, it reduces to a point density.

Building on this, the density of any unit norm vector, $\mathbf{x}_i$ drawn from a mixture of $Z$ different vMF distributions is defined as:

\begin{equation}
 \label{eq:moVMF}
  \rho(\mathbf{x}_i;{\bm\mu}_z\}^Z_{z=1},\{\kappa_z\}^Z_{z=1})=\sum^{K}_{z=1}\pi_z \rho(\mathbf{x}_i;\bm{\mu}_z,\kappa_z)
\end{equation}

where $\pi_z \geq 0$ and $\sum^{Z}_{z=1}\pi_z =1$. An Expectation-Maximization (EM) algorithm is used for estimating the mixture parameters. The mixture log-likelihood is defined as: 

\begin{equation}
\begin{aligned}
 \label{eq:mll}
l(\mathcal{X};\{\pi_z,\boldsymbol{\mu}_z,\kappa_z\}^Z_{z=1})=\sum\limits_{i=1}^N (\textrm{log} \sum\limits_{z=1}^Z\; \rho(\mathbf{x}_i;\boldsymbol{\mu}_z,\kappa_z))
\\
\text{subject to}\;\; \vert\vert \boldsymbol{\mu}_z\vert\vert=1 \;\; \textrm{and} \;\;\kappa_z \geq 0.
\end{aligned}
\end{equation}

In the E-step, the probability of $\mathbf{x}_i$ being sampled from $z$th mixture ($\gamma_{iz}$) is calculated as:

\begin{equation}
 \label{eq:gamma}
\gamma_{iz}=\frac{\pi_z\rho(\mathbf{x}_i;\bm{\mu}_z,\kappa_z)}{\sum\limits_{l=1}^{Z}\pi_l\rho(\mathbf{x}_i;\bm{\mu}_l,\kappa_l)}.
 \end{equation}

During the M-step of EM, this $\gamma_{iz}$ is used to calculate parameter updates which maximizes the log-likelihood defined in equation \ref{eq:mll}. These parameter updates are defined below:

\begin{equation}
\label{eq:em_update}
\begin{aligned}
\pi_z&=\frac{1}{N} \sum\limits_{i=1}^N \gamma_{iz},
\;\;\boldsymbol{\mu}_z=\frac{1}{N} \sum\limits_{i=1}^N \mathbf{x}_i\gamma_{iz}, \;\;
\overline{r}=\frac{\Vert \boldsymbol{\mu}_z \Vert}{N\pi_z},\\
\boldsymbol{\mu}_z&=\boldsymbol{\mu}_z/\Vert\boldsymbol{\mu}_z\Vert, \;\;
\kappa_z=\overline{r}(wd-\overline{r}^2)/(1-\overline{r}^2).
\end{aligned}
\end{equation}

The complete EM algorithm and detailed derivation of EM updates is documented in Banerjee \emph{et al} \cite{banerjee}.

\subsection{Obtaining directional embeddings (DE) using moVMF}
\label{ssec:DE}
The trained moVMF model is used to obtain directional embeddings (DE) of an input test audio recording. An under-complete dictionary, $\mathbf{M} \in \mathbb{R}^{wd\times Z}$ ($wd$ is dimensions of a super-frame and $Z$ is the number of unit mean vectors), is formed using the mean directions of the mixtures ($\boldsymbol{\mu}_z$) as its atoms. This dictionary forms the reference vocalization model. The mean directions of the mixtures having very low concentration are not included in the dictionary. To obtain the DE for an audio recording, its super-frame representation, $\mathbf{P} \in \mathbb{R}^{wd\times K}$, ($wd$ is dimensions of a super-frame and $K$ is the number of super-frames in the test audio recording) is projected on $\mathbf{M}$ as: $\mathbf{F_{proj}}=\mathbf{M}^T \times \mathbf{P}$, $\mathbf{F_{proj}}\in \mathbb{R}^{Z\times K}$ whose  columns contains the $Z$-dimensional DE representation for the input super-frames.

\begin{figure}[t]
	\centering
	\includegraphics[scale=0.5]{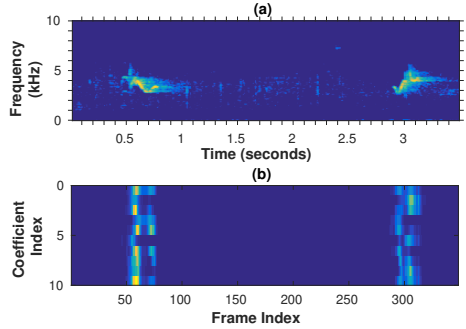}
	\caption{(a) Spectrogram containing two vocalizations of Cassin's Vireo (b) DE obtained using the proposed approach from (a).}
	\label{fig:feature1}
\end{figure}

\textbf{\emph{Nature of DE coefficients:}} The magnitude of DE coefficients ($\mathbf{F_{proj}}$) obtained from a super-frame corresponding to the bird vocalizations is significantly higher than the magnitude of DE coefficients calculated from the background regions. This can be attributed to the differences in alignments of bird and background super-frames with the unit mean directions (which form the atoms) accumulated in the dictionary. A bird vocalization super-frame is aligned in the directions of unit mean vectors, which leads to the high magnitude of DE coefficients. On the other hand, the background super-frames show no correlations with the dictionary atoms. Thus, the DE coefficients of background regions are characterized by their relatively low magnitude. This difference in magnitude of the DE coefficients leads to the distinction between bird vocalizations and background. This behavior is illustrated in Fig.~\ref{fig:feature1}. A dictionary (containing 10 mean directions) learned from the vocalizations of Cassin's vireo, a North American songbird, is used to obtain DE (shown in Fig.~\ref{fig:feature1}(b)) from the spectrogram shown in Fig.~\ref{fig:feature1}(a). A context window of 5 i.e. $w=5$ is used for obtaining the super-frames. By analyzing Fig.~\ref{fig:feature1}(b), it is clear that only bird vocalization information is reflected in DE space and most of the other background disturbances are implicitly removed. It must be noted that no explicit filtering process is applied in DE space to remove the background sounds. This background filtering is a result of non-alignment of background super-frames with the dictionary atoms.

 \begin{figure}[t]
	\centering
	\includegraphics[scale=0.5]{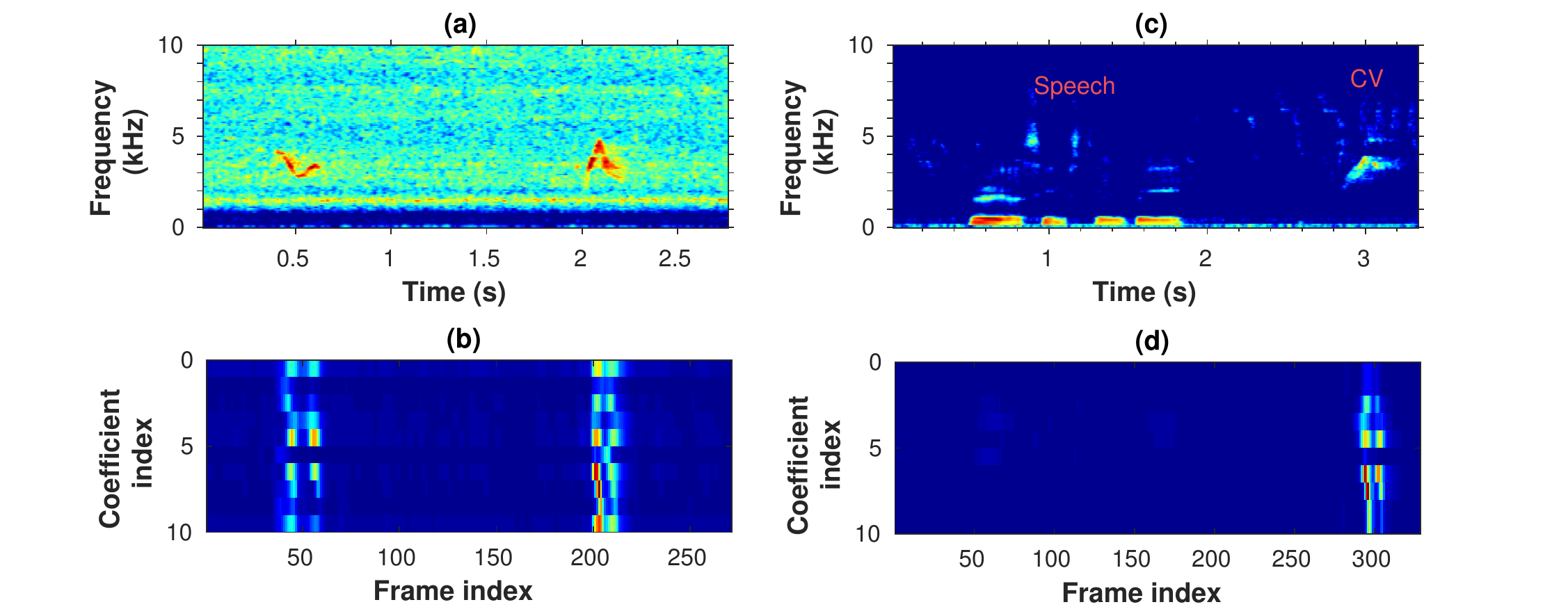}
	\caption{(a) and (c) depicts the spectrograms of recordings having vocalizations of Cassin's Vireo with overwhelming waterfall noise and human speech respectively. (b) and (d) illustrates their respective DE representation.}
	\label{fig:waterfall_human}
\end{figure}

\begin{figure}[t]
	\centering
	\includegraphics[scale=0.5]{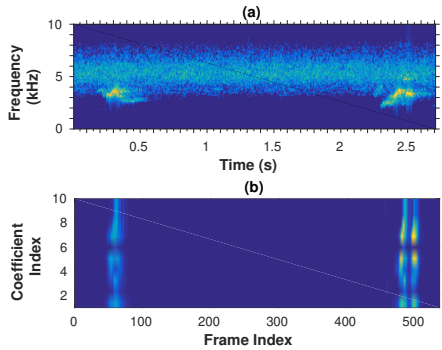}
	\caption{(a) The spectrogram of an audio recording having Cassin's Vireo vocalizations with significant background noise produced by cicadas (b) DE representation obtained from the spectrogram shown in (a).}
	\label{fig:cicada}
\end{figure}

\begin{figure}[t]
	\centering
	\includegraphics[scale=0.6]{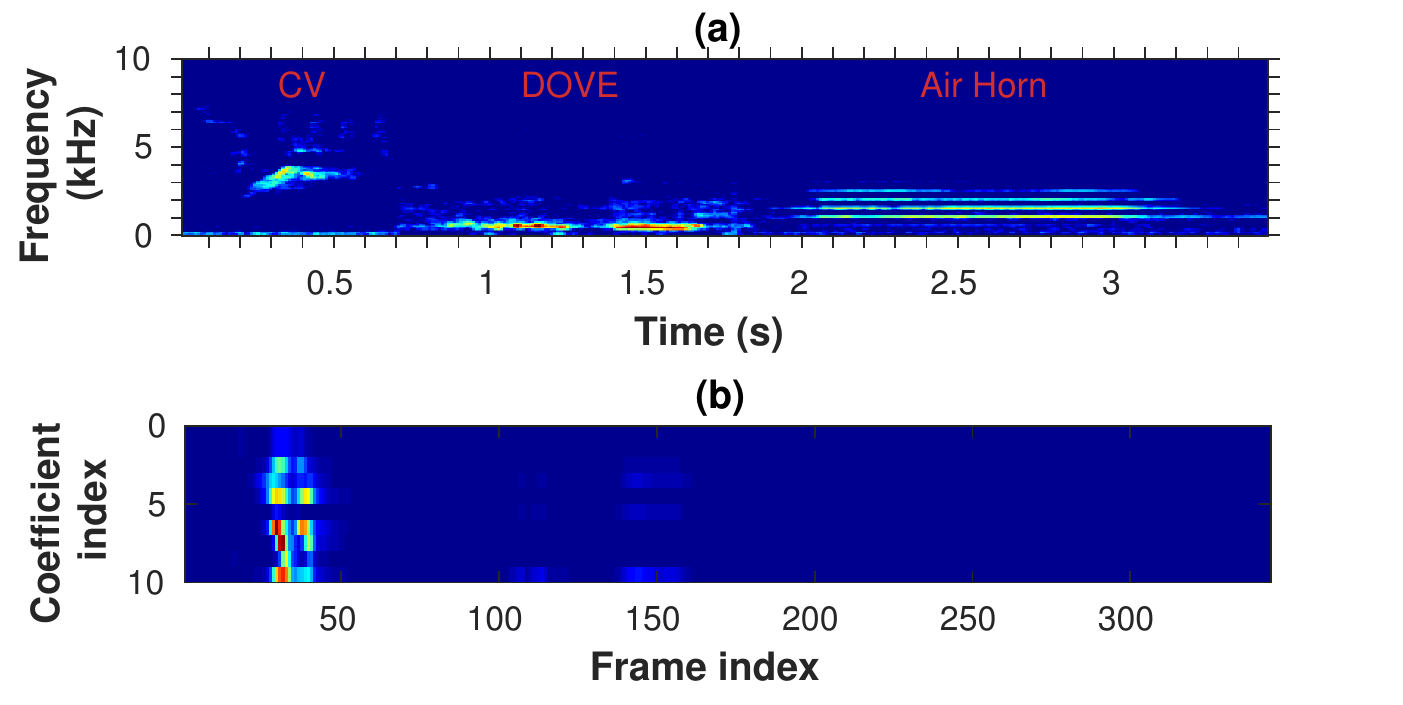}
	\caption{(a) Spectrogram containing a Cassin's vireo vocalization, a low frequency emerald dove vocalization and an air horn (b) DE obtained from the spectrogram shown above in (a).}
	\label{fig:siren_dove}
\end{figure}

Directional embeddings can also distinguish bird vocalizations from various types of biotic and abiotic background sounds such as the sound of rain, waterfall, human speech and insects. This behavior is highlighted in Figures \ref{fig:waterfall_human}, \ref{fig:cicada} and \ref{fig:siren_dove}. Fig.~\ref{fig:waterfall_human}(a) shows the spectrogram of two Cassin's vireo vocalizations with overwhelming waterfall sound present in the background (SNR 0 dB). The DE obtained for this spectrogram, depicted in Fig.~\ref{fig:waterfall_human}(b), indicates that the proposed representation is able to filter-out most of the background.  Fig.~\ref{fig:waterfall_human}(c) and \ref{fig:waterfall_human}(d) depict the spectrogram of an audio recording containing human speech with a Cassin's vireo vocalization and its respective DE. It is clear from Fig.~\ref{fig:waterfall_human}(d) that, as desired, the speech information is not reflected in the DE space. Also, Fig.~\ref{fig:cicada} depicts the robustness of DE against a commonly occurring biotic background disturbance in the field conditions, namely cicadas. These insects are highly vocal and often overwhelm the bird vocalizations, making it difficult to correctly segment them. Fig.~\ref{fig:cicada}(a) shows the spectrogram of an audio containing Cassin's vireo song phrases with cicadas constantly vocalizing in the background. DE representation obtained for this spectrogram is shown in Fig.~\ref{fig:cicada}(b). The analysis of this figure highlights that DE is correctly able to identify the bird vocalizations even in the presence of overwhelming noise produced by cicadas. The same dictionary as described earlier is also used here.

\begin{figure}[h]
	\centering
	\includegraphics[width=0.9\textwidth,height= 11 cm]{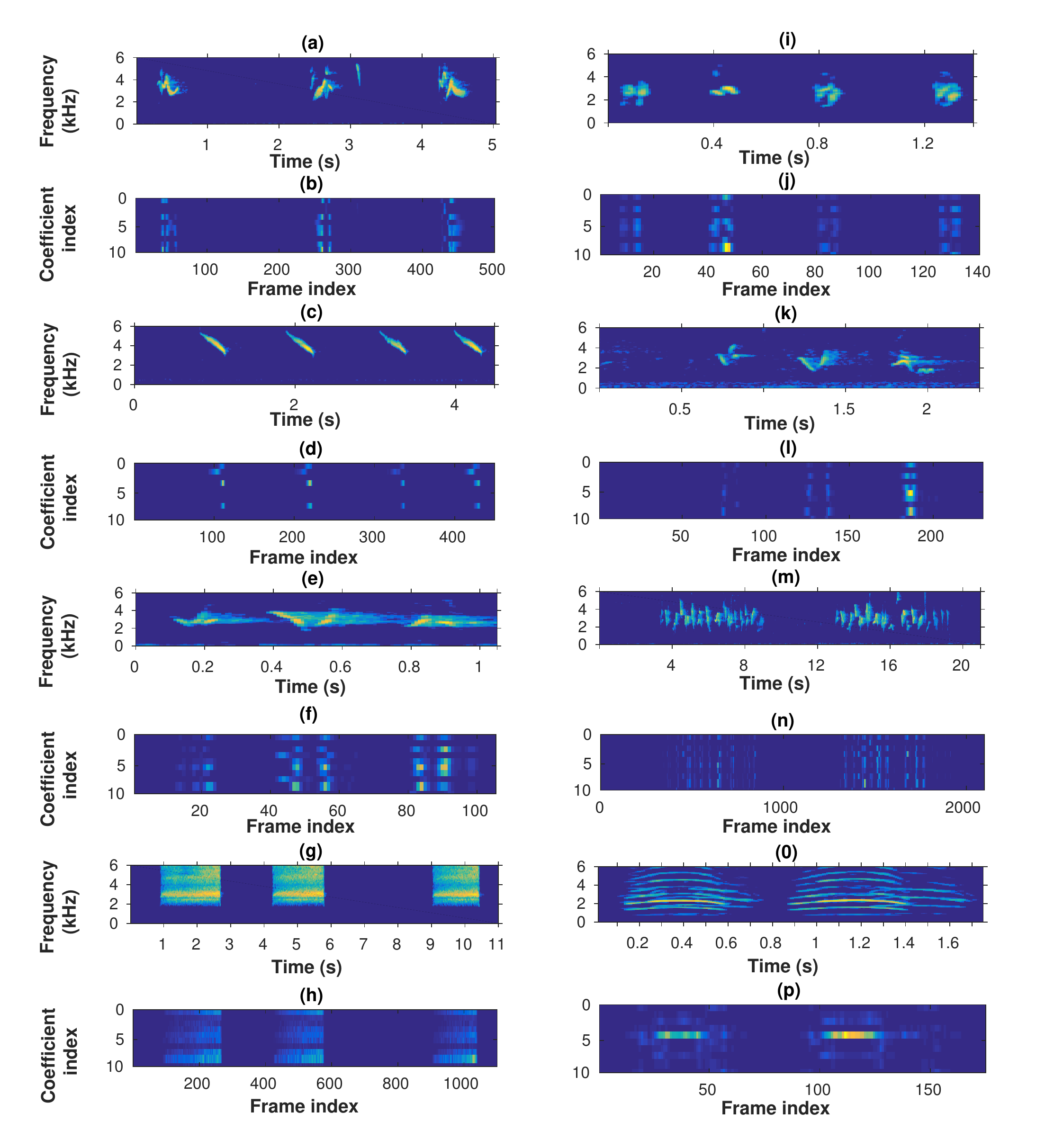}
	\caption{(a), (c), (e), (g), (i), (k), (m) and (O) depict the spectrograms containing vocalizations of Cassin's vireo, Hutton vireo, Black-headed grosbeak, Greater sooty owl, California thrasher, Western tanager, Black phoebe and Indian peafowl respectively. Their respective feature representations/DE are shown in (b), (d), (f), (h), (j), (l), (n) and (p).}
	\label{fig:multiple}
\end{figure}

To further analyze this behavior, we created an artificial audio recording containing a Cassin's vireo vocalization, an emerald dove vocalization (which has relatively low frequency vocalizations) and the sound of a siren. The spectrogram of this audio recording is shown in Fig.~\ref{fig:siren_dove}(a). Again, the aforementioned dictionary learned from Cassin's vireo is used. The analysis of Fig.~\ref{fig:siren_dove}(b) shows that the DE space has most of the information about the Cassin's vireo vocalization. However, no dominant information about emerald dove vocalization or the siren is present. This is due to the fact that mean directions are learned from the vocalizations of Cassin's vireo, whose frequency profile is significantly different from the frequency profiles of dove vocalizations and siren sounds. In many applications,  the siren is undesired (which is achieved in the DE space). Other applications might require that both the dove and Cassin's vireo vocalizations are captured properly (which is not achieved in this illustration.)  This exhibits the trade-off between the generalization ability and background removal ability of DE. The DE can be generic as long as the frequency profile of target vocalizations is similar to the ones used for learning the dictionary. This behavior may not an issue in many field conditions, as a dictionary learned from one species can be used to segment vocalizations of many target species, which exhibit similar frequency profiles. For example, a Cassin's vireo dictionary can be used to segment vocalizations of many song bird species. Fig.~\ref{fig:multiple} depicts the DE obtained for eight different species using Cassin's vireo dictionary. The analysis of this figure highlights the generalization claim (under restriction of the similarity between the reference and testing vocalizations). This claim is further corroborated in the experimentation (see sections \ref{sec:exp1} and \ref{sec:results}).                

\subsection{Comparison with Non-negative matrix factorization}
\begin{figure}[t]
	\centering
	
\includegraphics[trim={0cm 0cm 1cm 0cm},clip,scale=0.5]{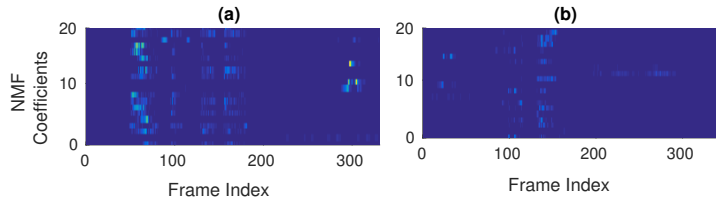}
	\caption{(a) and (b) depicts the feature representations of spectrograms shown in Figures \ref{fig:waterfall_human}(c) and \ref{fig:siren_dove}(a) obtained by using semi-supervised NMF (with 20 atoms).}
	\label{fig:NMF}
\end{figure}

Non-negative matrix factorization (NMF) has been successfully used in various tasks such as acoustic scene classification \citep{bisot}, speech enhancement \citep{nmf_enhancement} and voice activity detection \citep{germain}. For the task of bird vocalization segmentation, a dictionary or reference model of vocalizations can be learned using supervised or semi-supervised NMF \citep{nmf}. Since the proposed framework does not use the background modeling, we are only considering the semi-supervised NMF where the vocalization dictionary is kept constant and the background atoms are updated in each iteration. The feature representation obtained using NMF can distinguish between bird sounds and silence regions effectively. However, this representation also reflects the presence of most of the background sounds. Both of these behaviors are evident in Fig.~\ref{fig:NMF}. The semi-supervised NMF features (see Fig.~\ref{fig:NMF}) obtained for spectrograms shown in Fig.\ref{fig:waterfall_human}(c) and \ref{fig:siren_dove}(a) show the significant presence of background acoustic events in the feature domain. This can be attributed to the fact that NMF models the source spectra by a convex cone which often contains the regions where source spectra does not exist \cite{mld}. Moreover, these regions can contain other undesirable background acoustic events. This may hinder the ability of NMF to distinguish bird vocalizations from other background sounds in a semi-supervised setup.

In addition to this, obtaining NMF representation is computationally expensive (iterative and requires multiple matrix updates) in comparison to the calculation of DE from a pre-trained dictionary.                      

\section{Method}
\label{sec:proposed}
In this section, first, we formulate the problem statement and then, we describe the proposed framework for bird vocalization segmentation. 

\subsection{Problem formulation} 
The objective of the bird vocalization segmentation is to temporally locate any bird vocalization irrespective of its nature (call or song) and the species producing the vocalization. There are two steps in formulating this problem: feature representation and classification. First, a short-term feature representation is obtained from an input recording, where a feature vector $\mathbf{x}_i \in \mathbb{R}^y$ ($y$ is the number of features) is obtained for the $i$th frame. In classification stage, $\mathbf{x}_i$ is classified as the bird vocalization or the background. 


\subsection{Proposed framework}
The proposed two-pass framework (Fig.~\ref{fig:framework}) for bird vocalization segmentation is described in this sub-section. Here, first we describe pass 1 of the framework: the process to generate automatic training labels from an input test recording using DE and mutual information. Then, we discuss pass 2 of the framework, where labels and the DE features, generated during pass 1 from the input recording itself, are used for training a classification model which in turn is used for making bird/background decisions.  
\begin{figure}[t]
	\centering
	\includegraphics[scale=0.5]{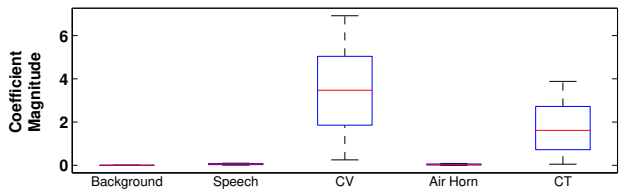}
	\caption{Box plots of the DE coefficients calculated for (a) 100 background frames (b) 100 speech frames and (c) 100 Cassin's vireo (CV) vocalization frames (d) 100 air-horn frames and (e) 100 Califorina Thrasher (CT), a bird species frames. In all cases, the dictionary/reference model learned from Cassin's vireo vocalizations is used.}
	\label{fig:magnitude}
\end{figure} 

\subsection{Pass 1: Generating automatic training labels}
\textbf{\emph{Obtaining DE from input audio recording}}:
A test audio recording and a dictionary containing unit mean directions of a trained moVMF are given as input to pass 1 of the framework. The input audio recording is converted into the super-frame based representation as described in Section \ref{ssec:prep}. This super-frame based representation is projected on the dictionary  to obtain the DE representation. This process is discussed in detail in Section \ref{ssec:DE}.

\textbf{\emph{Using mutual information to automatically generate labels}}:
As discussed in Section \ref{ssec:DE}, the magnitude of DE coefficients obtained from non-bird super-frames is significantly less than DE coefficients of bird vocalization super-frames. Also, there is a large variation in magnitudes of DE coefficients of the bird super-frames. This is because a bird super-frame exhibits varying degree of alignments with different unit mean directions of the dictionary. However, there is very little variation in magnitude of DE coefficients of a non-bird super-frame. This can be attributed to the fact that a non-bird super-frame has no or very little similarity with any of the dictionary atoms. This dissimilarity leads to low magnitudes (approaching zero) of DE coefficients. Fig.~\ref{fig:magnitude} highlights this behavior. 

\begin{figure}[h]
	\centering
	\includegraphics[trim={3cm 1cm 5cm 0cm},scale=0.65]{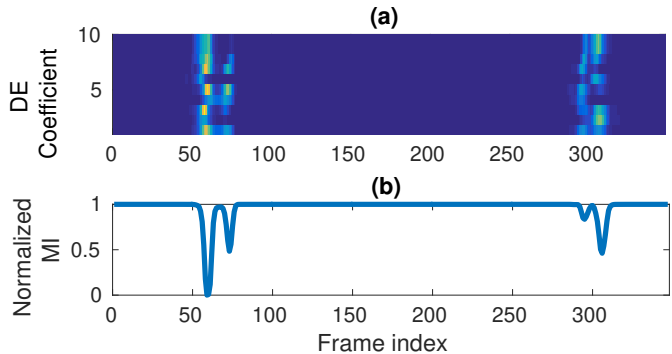}
	\caption{(a) DE representation obtained for spectrogram shown in Figure \ref{fig:waterfall_human}(a). (b) Normalized mutual information calculated for the DE representation shown in (a).}
	\label{fig:metric_clean}
\end{figure}

The difference in magnitudes of DE coefficients between bird and non-bird regions can be captured using mutual information (MI) between DE features of two consecutive super-frames. MI of a vector with itself is highest, therefore MI between two almost similar DE feature vectors will be higher than the two which are different. Since the DE feature vectors for any background sound are almost similar (all coefficients approach zero), MI calculated for background regions is almost constant and is higher than the MI calculated for the bird vocalization regions. Fig.~\ref{fig:metric_clean} illustrates this behavior.

To calculate MI, DE vectors ($\mathbf{F_{proj}}$) obtained from an input recording are normalized using the softmax function \cite{thakur_mlsp} as:
\begin{equation}
 \label{eq:MI}
 \mathbf{\hat{f}}_{n_j}=\frac{e^{\mathbf{f}_{n_j}}}{\sum_{z=1}^{Z}e^{\mathbf{f}_{n_z}}},\quad \textrm{for} \quad j=1,2,..,Z. 
\end{equation}
Here $\mathbf{f}_n$ is the $n$th column of $\mathbf{F_{proj}}$.  $\mathbf{f}_{n_j}$ represents the $j$th element of $\mathbf{f}_n$ and $Z$ represents the dimensionality of $\mathbf{f}_n$. Since softmax function calculates the categorical distribution of an event over $Z$ different events, $\mathbf{\hat{f}}_{n_j}$ represents the probability estimate of the degree of the alignment of $\mathbf{f}_n$ with $j$th mean direction of the learned dictionary $\mathbf{M}$. As discussed earlier, the DE vectors of non-bird regions exhibit no or very low alignment with any of the dictionary atoms. As a result, these non-bird DE vectors (softmax normalized) become almost equiprobable vectors i.e. each element of the normalized DE vector exhibits almost same value. Similarly, elements of a normalized DE vector, corresponding to a bird super-frame, exhibits different probabilities of the degree of alignment of super-frame with different mean directions.

MI between the normalized DE vectors of each pair of the consecutive super-frames (i.e. between $n_{th}$ and $(n-1)_{th}$ columns of normalized $\mathbf{F_{proj}}$) is calculated as:    
\begin{equation}
 \label{eq:MI}
 MI(\mathbf{\hat{f}}_n,\mathbf{\hat{f}}_{n-1}) = H(\mathbf{\hat{f}}_n) + H(\mathbf{\hat{f}}_{n-1}) - H(\mathbf{\hat{f}}_n,\mathbf{\hat{f}}_{n-1}).
\end{equation}
Here $H()$ represents the entropy and can be calculated as:
 \begin{equation}
 H(\mathbf{\hat{f}}_n)=-\sum_{j=1}^{Z}\mathbf{\hat{f}}_{n_j}\log_2\mathbf{\hat{f}}_{n_j},\;\; \textrm{where Z is the dimensions of DE vector}.
 \end{equation}
Also, $\mathbf{\hat{f}}_{n_j}$ is a multinoulli probability estimate (as discussed earlier) and $Z$ is the dimensions of $\mathbf{\hat{f}}_n$. 

The joint entropy $H(\mathbf{\hat{f}}_n,\mathbf{\hat{f}}_{n-1})$ can be calculated as:
 \begin{equation} 
 H(\mathbf{\hat{f}}_{n-1},\mathbf{\hat{f}}_n)=-\sum_{j=1}^{Z}P(\mathbf{\hat{f}}_{n-1_j}, \mathbf{\hat{f}}_{n_j}) \log_2P(\mathbf{\hat{f}}_{n-1_j},\mathbf{\hat{f}}_{n_j}),
 \end{equation}
where the joint probability estimate $P(\mathbf{\hat{f}}_{n-1_j}, \mathbf{\hat{f}}_{n_j})$ is calculated as:
\begin{equation}
P(\mathbf{\hat{f}}_{n-1_j},\mathbf{\hat{f}}_{n_j})=\frac{\textrm{number of times}\; (\mathbf{\hat{f}}_{n-1_j}, \; \mathbf{\hat{f}}_{n_j})\; \textrm{jointly occurs in}\;\mathbf{\hat{f}}_{n-1}\; \textrm{and}\; \mathbf{\hat{f}}_{n}}{Z}    
\end{equation}

MI, calculated using equation \ref{eq:MI}, is used to generate automatic training labels in the proposed framework. $Q$ super-frames having lowest and $Q$ super-frames having highest MI are labeled as bird vocalizations and background respectively. The rest of the super-frames remain unused. These generated labels and DE feature vectors are given as input to the second pass.

\subsection{Pass 2: Training classification model and extracting segments}
During the second pass, a classification model is trained using the input labels (generated during pass 1) and DE feature vectors. Once the model is trained, each DE vector of the input recording is classified as either the background or bird vocalization. In this work, we have used Support vector machines (SVM) for classification. Since the features generated from the input recording is itself used for training the classification model, the proposed framework does not suffer from the classical problem of mismatch in training-testing conditions. In bioacoustics, this mismatch is very common due to the volatile/unpredictable nature of the field conditions and differences in the audio recording devices.

\begin{figure}[t]
	\centering
	\includegraphics[trim={7cm 1cm 7cm 1cm},scale=0.65]{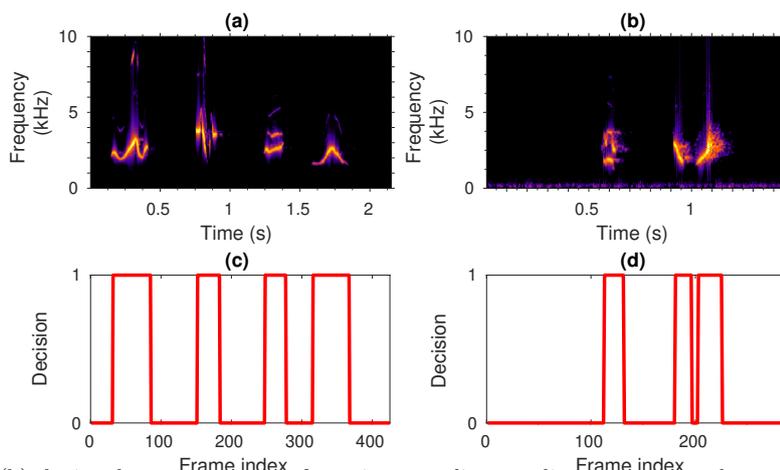}
	\caption{(a) and (b) depict the spectrograms of two input audio recordings containing the vocalizations of Black-headed grosbeak and California thrasher respectively. (c) and (d) show the decisions outputted by the proposed framework on these audio recordings.}
	\label{fig:results}
\end{figure}

Fig.~\ref{fig:results}(c) and Fig.~\ref{fig:results}(d) show the output of the proposed framework for the input test audio recordings whose spectrograms are shown in Fig.~\ref{fig:results}(a) and Fig.~\ref{fig:results}(b) respectively. The `0' represents the background while the `1' represents the bird vocalization. The analysis of this figure makes it clear that the proposed framework can identify the bird vocalizations in an audio recording successfully.

\section{Experiments}
\label{sec:exp1}
This section describes the datasets and experimental setup used for the performance evaluation of the proposed segmentation framework.

\subsection{Datasets Used}
Experimental validation of the proposed framework is performed on the audio recordings of seven different species. These species include Cassin's Vireo (CV1), California Thrasher (CT), Hutton Vireo (HV), Black-headed Grosbeak (BG), Grey Shrike-Thrush (GS), Redthroat (RT) and  Western Tanager (WT). The recordings of these 7 bird species were obtained from the bird audio database maintained by the Art \& Science center, UCLA\cite{data2}. Apart from this, a different set of Cassin's Vireo recordings (CV2)\cite{data1} is also used for the performance evaluation. All the audio recordings available at both these sources are 16-bit mono WAV files having a sampling rate of 44.1 kHz. The details about the number of recordings and the number of vocalizations available for each species are tabulated in Table \ref{dataset}. The choice of species used in this study is restricted by the public availability of the strongly labeled datasets (having segment level time stamps). The datasets used in this study have both labeled and unlabeled audio recordings. However, only the labeled audio recordings from all datasets are used here. A list of audio file used in this study and the process to retrieve these files is available at \url{https://bit.ly/2Oth03R}. The ground truth provided with datasets is in terms of time-stamps (onset/offset times) of the bird vocalizations. These time-stamps are processed to obtain frame-level ground truths, which are used for evaluating the segmentation performance.

To test the proposed framework in noisy conditions, background noise are artificially added to the recordings of the CV2 dataset. Four different types of background sounds i.e. rain, waterfall, river and cicadas at 0 dB, 5 dB, 10 dB, 15 dB and 20 dB SNR are added using Filtering and Noise Adding Tool (FaNt) \cite{fant}. These background sound files are downloaded from FreeSound \cite{freesound}. The bird vocalization recordings are most likely to be affected by similar sounds in the field conditions. 

\begin{table}[t]
\centering
\caption{Total audio recordings and number of vocalizations present in each dataset considered for the performance evaluation.}

\label{dataset}
\resizebox{1\textwidth}{!}{ 
\begin{tabular}{c|c|c|c|c|c|c|c|c|c}
\hline \hline
\textbf{Dataset}                                                           & CV1 & CV2    & CT    & HV  & BG  & GS & RT  & WT &Total  \\ \hline
\textbf{\begin{tabular}[c]{@{}c@{}}No. of Recordings\end{tabular}}        & 497  & 13   & 97    & 5   & 108   & 12  & 7    & 3  &744 \\ \hline
\textbf{\begin{tabular}[c]{@{}c@{}}Vocalizations (Approx.)\end{tabular}} & 40K & 900 & 20K & 278 & 16K & 120 & 1.2K & 500 &78998 \\ \hline \hline 
\end{tabular}
}
\end{table}

\subsection{Performance Evaluation Setup}
Three experiments are designed to evaluate the segmentation performance of the proposed framework. In the first experiment, the segmentation performance of the proposed framework is compared with various existing algorithms over the datasets mentioned earlier. Also, using the noisy versions of CV2, the robustness of the proposed framework and other comparative algorithms is evaluated in different SNR conditions. The second experiment is conducted to highlight the generic nature of the proposed framework. In this experiment, the dictionary learned from one species is used to segment the vocalizations of another species. The performance of this setup is compared with the segmentation performance obtained when same species is involved in both training and testing. For performance comparison, frame-level F1-score is used as a performance metric.  

For comparison, six different existing methods (discussed in section \ref{sec:background}) are used as baselines in this study. Apart from these algorithms, a variant of the proposed framework is also used for comparative studies. This variant targets the feature representation module of the proposed framework by replacing the DE with a semi-supervised non-negative matrix factorization (NMF) (discussed in section \ref{sec:features_learning}). Table \ref{algos} summarizes all the algorithms used for comparative study. 
%

Third experiment is conducted to compare the performances of the weakly supervised neural networks with the proposed framework. In section \ref{sec:1}, the authors explained that the problem of the scarcity of strong training labels for the task in hand can also be addressed by the data-intensive, weakly supervised neural networks that are trained for audio tagging tasks. These networks only require the recording level labels during training and bypass the need of strong or frame-level labels. The frame-level predictions generated by these networks are the by-product of audio-tagging tasks. Although, to the best of our knowledge, no existing study has used weakly supervised neural networks for bird vocalization segmentation. However, for the proper performance evaluation of the proposed framework, it is essential to incorporate such weakly supervised neural networks in the performance comparison. Thus, we utilize \emph{off-the-shelf} weakly supervised convolutional neural networks, proposed in \cite{large_scale,xu_attention} for the performance comparison\footnote{Python codes available at:\\\url{https://github.com/yongxuUSTC/dcase2017_task4_cvssp/} \\\url{https://github.com/yongxuUSTC/att_loc_cgrnn/}}. In \cite{xu_attention}, the authors used attention and localization mechanism in a CNN-RNN (convolutional-recurrent neural network) architecture (\emph{Att\_loc\_CNN}) to detect the audio events. In \cite{large_scale}, a gated convolution neural network (\emph{GCNN}) is proposed for detecting rare sound events. The gated convolutional layers control the flow of information from layer to other and act as attention mechanism. Both these networks can be easily extended for bird activity detection (audio tagging task) and hence, bird vocalization segmentation. Since the nature of methods listed in Table \ref{algos} is completely different from the weakly supervised neural networks, it is not possible to use the same type and amount of data for training. Hence, we train these networks on audio recordings obtained from $Wablr$ and $Freefield$ datasets. These datasets are provided for training as a part of \emph{Bird Activity Detection Challenge, 2017} \cite{bad}. A total 16000 recordings (8000 bird and 8000 non-bird approx.) are used for training both the networks. Each recording is 10 seconds long and is sampled at 44.1 kHz. The testing is done on all the datasets listed in Table \ref{dataset}.

\begin{table*}[t]
\centering
\caption{Methods used for comparative studies}
\label{algos}
 \resizebox{\textwidth}{!}{ 
\begin{tabular}{c|c|c}
\hline \hline
\textbf{Algorithm}                                                                                                    & \textbf{Classification Method} & \textbf{Nature} \\ \hline

Two-pass segmentation (\textbf{\textit{TP}}) \citep{thakur2016}                                                                                           & Gaussian Mixture Models (GMM)  & Unsupervised    \\ \hline
SVD + MI (\textbf{\textit{SVD}}) \citep{thakur_mlsp}                                                                                                       & Thresholding                   & Semi-supervised \\ \hline

Morphological noise removal on spectrograms + Energy (\textbf{\textit{FOD}}) \citep{fodor}                                                                   & Thresholding                   & Unsupervised    \\ \hline
Morphological noise removal and finding connected segments  + Energy (\textbf{\textit{LAS}}) \citep{las}                                                                   & Thresholding                   & Unsupervised    \\ \hline
Morphological Opened Spectrogram + Energy (\textbf{\textit{MOS}}) \cite{Oliveira}                                                                       & Thresholding                   & Unsupervised    \\ \hline
Spectrogram Enhancement and Directional Filtering + Energy (\textbf{\textit{DF}})  \citep{koluguri}                                                     & Thresholding                   & Unsupervised    \\ \hline
DE+SVM Based Framework (\textbf{\textit{Proposed}})                                                                                       & SVM                            & Semi-supervised \\ \hline
NMF+SVM (\textbf{\textit{NMF}}), variant of the proposed framework                                                                                                        & SVM                            & Semi-supervised \\ \hline \hline

\end{tabular}}
\end{table*}

\subsection{Parameter settings used}
\subsubsection{Parameters used in the proposed framework:}
For obtaining the spectrograms from audio recordings, STFT is performed on 20 ms frames with an overlap of 50\%, using 1024 FFT points. The super-frames are obtained by concatenating 5 frames i.e. $w=5$. The dictionary, $\mathbf{M}$, with 10 atoms is learned from the training vocalizations. moVMF with 15 mixtures is applied to obtain the dominant directions of bird vocalizations. Out of these 15 mixtures, five having lowest concentration were discarded ($Z=10$). While auto-labeling super-frames in pass 1, 2000 super-frames having lowest MI are labeled as bird vocalizations while 2000 super-frames having highest MI are labeled as background (i.e $Q$=2000.) All these values are experimentally determined on a validation dataset. Support vector machines with cubic polynomial kernel is used for classification in pass 2. 

\subsubsection{Parameters used in the other comparative algorithms:}
For \textbf{\textit{SVD}}, the number of dictionary atoms used in the respective study \cite{thakur_mlsp} i.e. 5, is also used here. For implementing directional filtering (\textbf{\textit{DF}}), an order of 3 and a matrix length of 21 is used to apply Savitzky-Golay filtering. For \textbf{\textit{NMF}}, 20 dictionary atoms are used. A classical NMF algorithm \cite{lee2001algorithms} with Itakura-Saito distance is used in this work. All these parameters including thresholds are tuned empirically on a validation dataset to provide the optimal segmentation performance. For weakly supervised neural networks, only a small change is done in architectures to incorporate the required inputs. Mel-spectrograms are given as input to these networks. The input size is changed to $40 \times 1000$ to incorporate 10 seconds of audio recordings (1000 20 ms frames with 50\% overlap and 40 is the number of  Mel bands). The rest of parameters and architecture details proposed in the corresponding studies \cite{large_scale,xu_attention} are also used here.

\subsubsection{Data distribution for training, testing and validation:}
\textbf{\emph{For first experiment:}} From each dataset, two recordings are used as validation set for tuning the threshold and parameters. These recordings are not used for training or testing purposes. The remaining audio recordings are used for the evaluation in a five fold cross-validation manner and the results presented here are averaged across these five folds. For each fold, 20\% of the recordings are chosen from the dataset for learning the dictionary and rest of the recordings are used for testing. The same test recordings are given to the unsupervised methods for a fair performance comparison. It must be noted that the test recordings contain many vocalization types, which are not present in the training recording used for learning the dictionary.

\textbf{\emph{For second experiment (cross-species segmentation):}}  In this experiment, one dataset (corresponding to a particular species) is used for training i.e. for learning the dictionary while all the other datasets are used for testing. From the training dataset, two recordings are randomly chosen as the validation set and out of the remaining recordings, 20\% of recordings are randomly chosen for learning the dictionary. All the recordings present in the test datasets are used for evaluating the segmentation performance.

\textbf{\emph{For third experiment (comparison against weakly supervised neural networks):}}
For weakly supervised neural networks, 80\% of the recordings in \emph{Freefield} and \emph{Warblr} are used for training while remaining recordings are used for validation. The testing is done on all the datasets listed in Table \ref{dataset}. For a fair comparison, only the files used for testing the proposed framework in experiment 1 are used i.e. audio files used for validation are not used here. Also, the segmentation performances of these networks are compared against the average F1-score achieved across five-folds by the proposed framework.

\section{Results and Discussion}
\label{sec:results}
In this section, we describe the results obtained for three experiments mentioned in the previous section. In addition to that, we also discuss the various aspects of the proposed framework such as the effect of increasing the size of context window, increasing the number of mixtures in moVMF and using MI over energy/entropy.  

 \begin{table}[t]
\centering
\caption{Segmentation performance of different algorithms over different datasets (\emph{Results of first experiment}). F-Score is used as the metric for comparison. The dictionary is learned from the vocalizations of the species, which are targeted for segmentation. Entries in bold represent the best F-score obtained for the corresponding dataset.}
\label{exp1}
\begin{tabular}{c|c|c|c|c|c|c|c|c}
\hline \hline
\multirow{2}{*}{\textbf{Method}} & \multicolumn{8}{c}{\textbf{Datasets}}                                                                                      \\ \cline{2-9} 
                                 & \textbf{CV1}  & \textbf{CV2}  & \textbf{CT}   & \textbf{HV}   & \textbf{BG} & \textbf{GS}  & \textbf{RT} & \textbf{WT}   \\ \hline

TP                               & 0.69         & 0.72          & 0.68          & 0.7          & 0.64          & 0.64          & 0.66         & 0.69          \\ \hline
SVD                              & 0.79         & 0.82          & 0.77          & 0.8           & 0.76          & 0.78 & 0.79         & 0.8           \\ \hline

FOD                              & 0.61          & 0.62          & 0.58          & 0.67          & 0.59          & 0.6           & 0.63         & 0.65          \\ \hline
LAS                              & 0.6         & 0.6          & 0.57          & 0.68          & 0.57          & 0.57          & 0.61         & 0.64          \\ \hline
MOS                              & 0.52         & 0.57          & 0.53          & 0.65          & 0.57          & 0.58          & 0.6         & 0.62          \\ \hline

DF                               & 0.66         & 0.67          & 0.66          & 0.7           & 0.64          & 0.65          & 0.7         & 0.71           \\ \hline
Proposed                            & \textbf{0.81} & \textbf{0.83} & \textbf{0.8} & \textbf{0.8} & \textbf{0.77} & \textbf{0.79} & \textbf{0.81} & \textbf{0.81} \\ \hline
NMF                             & 0.76         & 0.78          & 0.76          & 0.76           & 0.74          & 0.72 & 0.72         & 0.76           \\ \hline \hline
\end{tabular}
\end{table}

\subsection{First Experiment: Comparison with existing methods}
The segmentation performance of the comparative algorithms including the proposed framework is depicted in Table \ref{exp1}. From the analysis of Table \ref{exp1}, the following can be inferred:

\begin{itemize}[leftmargin=*]
\item The proposed framework outperforms most of the existing algorithms across all the datasets considered. 
\item  \textbf{\textit{SVD}} and the proposed framework significantly outperform the unsupervised methods. This is due to their ability to discriminate bird vocalizations from most of the background disturbances.

\item \textbf{\textit{DF}} outperforms all the other spectrogram-processing based methods. This can be attributed to the spectrogram enhancement and directional filtering steps before applying energy-based thresholding. This processing on spectrogram significantly increases the energy contrast between vocalizations and background leading to a better segmentation. However, like other unsupervised methods, \textbf{\textit{DF}} is also not able to discriminate between bird and non-bird sounds.

\item NMF variant of the proposed framework also outperforms most of the existing algorithms and its segmentation performance is second only to the proposed framework and \textbf{\textit{SVD}}. Better segmentation performance can be attributed to the reference modeling of the bird vocalizations using NMF. However, as discussed earlier, NMF may not be able to discriminate between birds and other acoustic events effectively in a semi-supervised setup, which leads to a poorer segmentation performance than the proposed framework. 

\end{itemize}

    \begin{figure}[t]
	\centering
	\includegraphics[trim={0cm 1cm 8cm 0cm},clip,scale=0.6]{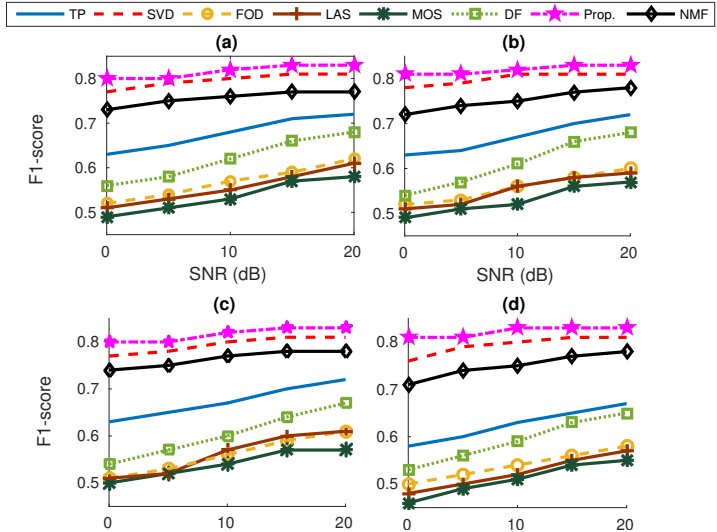}
	\caption{Segmentation performance of different algorithms on CV2 corrupted by (a) Rain, (b) River, (c) Waterfall and (d) Cicadas noise at different SNR. Here 'Prop.' represents the proposed framework.}  
	\label{fig:snr}
\end{figure}

The segmentation performance of these methods is also evaluated on four noisy versions of CV2, each having SNR of 0 dB, 5 dB, 10 dB, 15 dB and 20 dB.
The results are shown in Fig. \ref{fig:snr}. These results highlight the noise robustness of the proposed framework, which shows no significant drop in segmentation performance as the SNR deteriorates (20 dB to 0 dB). Apart from this, all the semi-supervised methods perform well at low SNR conditions. Again, the feature extraction step in all these methods mitigate the effect of the background noise to a large extent which results in better segmentation performance. On the other hand, all the unsupervised methods show a significant drop in performance on moving from high to low SNR conditions. An average relative drop of 12\%, 4.5\%,  16\%, 16.3\%, 15.5\%, 17\%, 3.6\% and 6.4\% is observed in performance of \textbf{\textit{TP}}, \textbf{\textit{SVD}}, \textbf{\textit{FOD}}, \textbf{\textit{LAS}}, \textbf{\textit{MOS}}, \textbf{\textit{DF}}, \textbf{\textit{Proposed}} and \textbf{\textit{NMF}} respectively on moving from 20 dB SNR to 0 dB SNR across all noise types.           
\begin{table}[h]
\centering
\caption{Segmentation performance for cross-species train-test setup (\emph{Results of the second experiment}). F-Score is used as the performance metric. The dictionary learned from one species is used for segmenting all the other species considered in the study. Entries in bold represent the best F-score obtained for the corresponding dataset.}
\label{generic}
\begin{tabular}{c|c|c|c|c|c|c|l}
\hline \hline
\multirow{2}{*}{\textbf{\begin{tabular}[c]{@{}c@{}}Training Dataset\end{tabular}}} & \multicolumn{7}{c}{\textbf{Testing Datasets}}                                                              \\ \cline{2-8} 
                                                                                     & \textbf{CT}   & \textbf{HV}   & \textbf{BG} & \textbf{GS}  & \textbf{RT} & \textbf{WT}   & \textbf{CV1}  \\ \hline
CT                                                                                   & \textbf{0.79} & \textbf{0.81} & \textbf{0.78} & 0.76          & 0.79         & 0.79          & 0.78         \\ \hline
HV                                                                                   & 0.77          & \textbf{0.81} & 0.76          & 0.76          & 0.78         & 0.79          & 0.77         \\ \hline
BG                                                                                 & 0.78          & 0.8           & \textbf{0.78} & \textbf{0.77} & 0.79         & 0.8           & 0.79         \\ \hline
GS                                                                                  & 0.76          & 0.8           & 0.77          & \textbf{0.77} & 0.78         & 0.78          & 0.78         \\ \hline
RT                                                                                  & 0.78          & 0.79          & 0.76          & 0.75          & \textbf{0.8} & 0.78          & 0.79         \\ \hline
WT                                                                                   & 0.77          & 0.8           & 0.77          & 0.75          & 0.79         & \textbf{0.81} & 0.78         \\ \hline
CV1                                                                                   & 0.78          & 0.8           & 0.77          & 0.76          & \textbf{0.8} & 0.8           & \textbf{0.8} \\ \hline \hline
\end{tabular}
\end{table}

\subsection{Second Experiment: Generic nature of the proposed framework}
In this experiment, the generic nature of the proposed framework is evaluated using different species for training (learning the dictionary) and testing. Table \ref{generic} tabulates the segmentation performance observed for this cross-species setup. The analysis of Table \ref{generic} shows that the best segmentation performance is obtained when the training and testing species are same. However, there is no dramatic drop in the performance when the dictionary is learned from the vocalization of a species that is not being targeted for the segmentation. Moreover, this drop in F1-score is minute and segmentation performance across all the cases is near optimum. This upholds the generalization claim of the proposed framework.   
\begin{table}[t]
\caption{Segmentation performance of the proposed framework against weakly supervised neural networks.}
\label{weak_results}
\begin{tabular}{c|c|c|c|c|c|c|c|c}
\hline\hline
\multirow{2}{*}{\textbf{Method}} & \multicolumn{8}{c|}{\textbf{Datasets}}                                                                                     \\ \cline{2-9} 
                                 & \textbf{CV1} & \textbf{CV2}  & \textbf{CT}  & \textbf{HV}  & \textbf{BG}   & \textbf{GS}   & \textbf{RT}   & \textbf{WT}   \\ \hline
GCNN \cite{large_scale}                             & 0.59         & 0.62          & 0.63         & 0.65         & 0.6           & 0.58          & 0.59          & 0.61          \\ \hline
Att\_loc\_CNN \cite{xu_attention}                    & 0.72         & 0.73          & 0.75         & 0.74         & 0.71          & 0.75          & 0.73          & 0.76          \\ \hline
Proposed                         & \textbf{0.8} & \textbf{0.83} & \textbf{0.8} & \textbf{0.8} & \textbf{0.77} & \textbf{0.79} & \textbf{0.81} & \textbf{0.81} \\ \hline\hline
\end{tabular}

\end{table}
\subsection{Third Experiment: Comparison against weakly supervised neural networks}
The performances achieved by the proposed framework and the weakly supervised neural networks on different datasets are listed in Table \ref{weak_results}. The analysis of Table \ref{weak_results} shows that the segmentation performance of the proposed framework is better than the weakly supervised neural networks considered in this study. The performance of \emph{GCNN}\cite{large_scale} and \emph{Att\_loc\_CNN}\cite{xu_attention} is not up to the expected standards as these networks were designed for rare sound event detection \footnote{\url{http://www.cs.tut.fi/sgn/arg/dcase2017/challenge/task-rare-sound-event-detection}} and not for the bird vocalization segmentation. Moreover, the differences in training and test datasets may have also amounted to the average performance of these neural networks. It must be noted that both these networks were able to identify the presence and absence of bird vocalizations. However, the onset-offset boundaries detected by \emph{Att\_loc\_CNN} are much better than boundaries detected by \emph{GCNN}. This can be attributed to the fact that \emph{Att\_loc\_CNN} utilizes time-distributed convolutional operations on each input frame, while \emph{GCNN} utilizes two-dimensional convolutional operations at each layer. Due to these two-dimensional convolutional operations, the \emph{field-of-view} observed at each frame in the last convolution layer spans multiple input frames \cite{deep_learning_book}. This elevated span deters the ability of network to detect the correct onset-offset times. Also, the results obtained from this experiment confirm that the weak supervision provides a convenient way to tap into the massive potential of huge datasets with weak labels.

\subsection{Effect of w and Z on segmentation performance of the proposed framework}
The context window (of size $w$) controls the amount of spectral-temporal information to be incorporated in a super-frame. This helps in capturing frequency and temporal modulations present in most of the bird vocalizations. The smaller value of $w$ gives rise to a super-frame representation having less context information while using a large value of $w$ increases the dimensionality of super-frames which is often undesirable. The large values of $w$ can also affect the generic nature of the proposed algorithm. The mean directions learned from the super-frames having very high context information can get highly biased towards the vocalizations used for training. Therefore, an appropriate value of $w$ must be used. As discussed previously, we have used $w=5$ for all the experiments. This value of $w$ is determined experimentally and it balances the trade-off between preserving the context information and high dimensionality of the super-frames.       

The number of mixtures ($Z$) in moVMF directly controls the dimensionality of the feature representation. We intend to train moVMF with less amount of training data (usually a single audio recording). Hence, the number of mixtures should not be very large. To choose an appropriate value of $Z$, we experimented with $Z=15,20\;\textrm{and}\;40$. The behavior of DE for all these cases was same i.e. no improvement was observed by increasing $Z$ from 15 to 40. This shows that the proposed DE representations can be learned effectively using few mean directions only. Considering the dimensionality of DE representation and usage of less training data, we chose to use $Z=15$ for all the experiments.

\subsection{Using MI over energy/entropy}
Most of the information in DE space is about the bird vocalizations. However, DE space can also exhibit background information to some extent (as can be seen in Figure \ref{fig:waterfall_human}(b)). These background disturbances are slowly varying and for two consecutive super-frames, DE representations are almost same. Hence, MI for background regions will still be almost constant. However, metrics like energy and entropy are known to be susceptible to these disturbances, leading to a varying energy/entropy for background regions. This may lead to the poor automatic labeling in the proposed framework. Therefore, MI is used in the proposed framework over energy and entropy.

\section{Conclusion}
\label{Sec:con}
In this paper, we proposed a two pass semi-supervised framework for bird vocalization segmentation. A feature representation called directional embedding, tailored for this segmentation task, has also been proposed. The proposed framework gets its semi-supervised nature from the feature learning module which utilizes a small amount of training data to model the dominant directions of the vocalizations of many bird species. The utilization of this feature representation in the proposed framework helps in discriminating the bird vocalizations from other background acoustic events. The segmentation performance of the proposed framework has been evaluated on the datasets of seven different species and the experimental results establish the superiority of the proposed framework over various existing bird vocalization segmentation methods.
 
One possible shortcoming of the proposed framework is that when bird vocalizations used for learning the dictionary are similar to the undesired background events, the framework cannot discriminate between bird and non-bird sounds. However, it must be noted that these cases rarely occur in field conditions. Future work will deal with developing segmentation techniques which could tackle such situations. Also, the proposed DE representations will be explored in a supervised setup for species identification and acoustic event detection.


\section*{References}

\begin{thebibliography}{10}
\expandafter\ifx\csname url\endcsname\relax
  \def\url#1{\texttt{#1}}\fi
\expandafter\ifx\csname urlprefix\endcsname\relax\def\urlprefix{URL }\fi
\expandafter\ifx\csname href\endcsname\relax
  \def\href#1#2{#2} \def\path#1{#1}\fi

\bibitem{brandes2008automated}
T.~S. Brandes, Automated sound recording and analysis techniques for bird
  surveys and conservation, Bird Conservation International 18~(S1) (2008)
  S163--S173.

\bibitem{lee2008}
C.-H. Lee, C.-C. Han, C.-C. Chuang, Automatic classification of bird species
  from their sounds using two-dimensional cepstral coefficients, IEEE Trans.
  Audio, Speech, Language Process. 16~(8) (2008) 1541--1550.

\bibitem{fagerlund_thesis}
S.~Fagerlund, Automatic recognition of bird species by their sounds, Masters'
  thesis, Helsinki University Of Technology (2004).

\bibitem{dcase}
{DCASE} 2017, http://www.cs.tut.fi/sgn/arg/dcase2017/index, accessed:
  2018-04-05.

\bibitem{bad}
{B}ird {A}udio {D}etection {C}hallenge,
  \url{http://machine-listening.eecs.qmul.ac.uk/bird-audio-detection-challenge/},
  accessed: 2018-05-10.

\bibitem{large_scale}
Y.~Xu, Q.~Kong, W.~Wang, M.~Plumbley, Large-scale weakly supervised audio
  classification using gated convolutional neural network, in: Proceedings of
  Int. Conf. Acoust., Speech, Signal Process., 2018.

\bibitem{nmf}
S.~Mohammed, I.~Tashev, A statistical approach to semi-supervised speech
  enhancement with low-order non-negative matrix factorization, in: Proceedings
  of Int. Conf. Acoust., Speech, Signal Process., 2017, pp. 546--550.

\bibitem{mld}
M.~Kim, P.~Smaragdis, Mixtures of local dictionaries for unsupervised speech
  enhancement, IEEE Signal Process. Lett. 22~(3) (2015) 293--297.

\bibitem{banerjee}
A.~Banerjee, I.~S. Dhillon, J.~Ghosh, S.~Sra, Clustering on the unit
  hypersphere using von mises-fisher distributions, J. Mach. Learn. Res.
  6~(Sep) (2005) 1345--1382.

\bibitem{dan_skm}
D.~Stowell, M.~D. Plumbley, Automatic large-scale classification of bird sounds
  is strongly improved by unsupervised feature learning, PeerJ 2 (2014) e488.

\bibitem{multi_skm}
S.~Dieleman, B.~Schrauwen, Multiscale approaches to music audio feature
  learning, in: Proceedings of Int. Soc. Music Info. Retrieval, 2013, pp.
  116--121.

\bibitem{zhong}
S.~Zhong, J.~Ghosh, A comparative study of generative models for document
  clustering, in: Proceedings of Workshop on clustering high dimensional data
  and its applications in SIAM data mining conference, 2003.

\bibitem{Trifa}
V.~M. Trifa, A.~N.~G. Kirschel, C.~E. Taylor, E.~E. Vallejo, Automated species
  recognition of antbirds in a {M}exican rainforest using hidden {M}arkov
  models, J. Acoust. Soc. Am. 123~(4) (2008) 2424--2431.
\newblock \href {http://dx.doi.org/http://dx.doi.org/10.1121/1.2839017}
  {\path{doi:http://dx.doi.org/10.1121/1.2839017}}.

\bibitem{Harma}
A.~Harma, P.~Somervuo, Classification of the harmonic structure in bird
  vocalization, in: Proceedings of Int. Conf. Acoust. Speech, Signal Process,
  2004, pp. 701--704.
\newblock \href {http://dx.doi.org/10.1109/ICASSP.2004.1327207}
  {\path{doi:10.1109/ICASSP.2004.1327207}}.

\bibitem{Somervuo}
P.~Somervuo, A.~Harma, S.~Fagerlund, Parametric representations of bird sounds
  for automatic species recognition, IEEE Trans. Audio, Speech, Language
  Process 14~(6) (2006) 2252--2263.
\newblock \href {http://dx.doi.org/10.1109/TASL.2006.872624}
  {\path{doi:10.1109/TASL.2006.872624}}.

\bibitem{Fagerlund}
S.~Fagerlund, Bird species recognition using support vector machines, EURASIP
  J. Appl. Signal Process. 2007~(1) (2007) 64--64.
\newblock \href {http://dx.doi.org/10.1155/2007/38637}
  {\path{doi:10.1155/2007/38637}}.

\bibitem{wang2013}
N.~C. Wang, R.~E. Hudson, L.~N. Tan, C.~E. Taylor, A.~Alwan, K.~Yao, Bird
  phrase segmentation by entropy-driven change point detection, in: Proceedings
  of Int. Conf. Acoust. Speech, Signal Process, 2013, pp. 773--777.
\newblock \href {http://dx.doi.org/10.1109/ICASSP.2013.6637753}
  {\path{doi:10.1109/ICASSP.2013.6637753}}.

\bibitem{Lakshmi}
B.~Lakshminarayanan, R.~Raich, X.~Fern, A syllable-level probabilistic
  framework for bird species identification, in: Proceedings of Int. Conf.
  Mach. Learn. Applicat., 2009, pp. 53--59.
\newblock \href {http://dx.doi.org/10.1109/ICMLA.2009.79}
  {\path{doi:10.1109/ICMLA.2009.79}}.

\bibitem{thakur2016}
A.~Thakur, P.~Rajan, Model-based unsupervised segmentation of birdcalls from
  field recordings, in: Proceedings of Int. Conf. Signal Process. Commun.
  Syst., 2016.

\bibitem{Oliveira}
A.~G. de~Oliveira, T.~M. Ventura, T.~D. Ganchev, J.~M. de~Figueiredo, O.~Jahn,
  M.~I. Marques, K.-L. Schuchmann, Bird acoustic activity detection based on
  morphological filtering of the spectrogram, Applied Acoustics 98 (2015)
  34--42.

\bibitem{fodor}
G.~Fodor, The ninth annual mlsp competition: first place, in: Proceedings of
  MLSP, 2013, pp. 1--2.

\bibitem{las}
M.~Lasseck, Large-scale identification of birds in audio recordings., in: CLEF
  (Working Notes), 2014, pp. 643--653.

\bibitem{glotin}
H.~Glotin, Y.~LeCun, T.~Artieres, S.~Mallat, O.~Tchernichovski, X.~Halkias, The
  {NIPS4B} 2013 competition, in: Proceedings of Neural Information Processing
  Scaled for Bioacoustics: from Neurons to Big Data, 2013.

\bibitem{koluguri}
N.~R. Koluguri, G.~N. Meenakshi, P.~K. Ghosh, Spectrogram enhancement using
  multiple window savitzky-golay (mwsg) filter for robust bird sound detection,
  IEEE/ACM IEEE Trans. Audio, Speech, Language Process. 25~(6) (2017)
  1183--1192.

\bibitem{templates}
K.~Kaewtip, L.~N. Tan, C.~E. Taylor, A.~Alwan, Bird-phrase segmentation and
  verification: A noise-robust template-based approach, in: Proceedings of Int.
  Conf. Acoust. Speech, Signal Process, 2015, pp. 758--762.

\bibitem{Neal}
L.~Neal, F.~Briggs, R.~Raich, X.~Z. Fern, Time-frequency segmentation of bird
  song in noisy acoustic environments, in: Proceedings of Int. Conf. Acoust.
  Speech, Signal Process, 2011, pp. 2012--2015.
\newblock \href {http://dx.doi.org/10.1109/ICASSP.2011.5946906}
  {\path{doi:10.1109/ICASSP.2011.5946906}}.

\bibitem{thakur_ncc}
A.~Thakur, P.~Rajan, Unsupervised birdcall activity detection using source and
  system features, in: Proceedings of NCC, IEEE, 2017.

\bibitem{thakur_mlsp}
A.~Thakur, P.~Rajan, R{\'e}nyi entropy based mutual information for
  semi-supervised bird vocalization segmentation, in: Proceedings of MLSP,
  2017.

\bibitem{marler}
P.~Marler, Bird calls: their potential for behavioral neurobiology, Annals of
  the New York Academy of Sciences 1016~(1) (2004) 31--44.

\bibitem{tang2009}
H.~Tang, S.~M. Chu, T.~S. Huang, Generative model-based speaker clustering via
  mixture of von mises-fisher distributions, in: Proceedings of Int. Conf.
  Acoust. Speech, Signal Process, 2009, pp. 4101--4104.

\bibitem{bangert}
M.~Bangert, P.~Hennig, U.~Oelfke, Using an infinite von mises-fisher mixture
  model to cluster treatment beam directions in external radiation therapy, in:
  Proceedings of Int. Conf. Mach. Learn. App., 2010, pp. 746--751.

\bibitem{text}
A.~N. Srivastava, M.~Sahami, Text mining: Classification, clustering, and
  applications, CRC Press, 2009.

\bibitem{shadow_cvpr}
A.~Panagopoulos, D.~Samaras, N.~Paragios, Robust shadow and illumination
  estimation using a mixture model, in: Proceedings of Comp. Vision and Pattern
  Recog., IEEE, 2009, pp. 651--658.

\bibitem{gopal}
S.~Gopal, Y.~Yang, Von mises-fisher clustering models, in: Proceedings of Int.
  Conf. on Machine Learn., 2014, pp. 154--162.

\bibitem{bisot}
V.~Bisot, R.~Serizel, S.~Essid, G.~Richard, Acoustic scene classification with
  matrix factorization for unsupervised feature learning, in: Proceedings of
  Int. Conf. Acoust., Speech, Signal Process., 2016, pp. 6445--6449.

\bibitem{nmf_enhancement}
N.~Mohammadiha, P.~Smaragdis, A.~Leijon, Supervised and unsupervised speech
  enhancement using nonnegative matrix factorization, IEEE Trans. Audio,
  Speech, Language Process. 21~(10) (2013) 2140--2151.

\bibitem{germain}
F.~G. Germain, D.~L. Sun, G.~J. Mysore, Speaker and noise independent voice
  activity detection, in: Proceedings of Interspeech, 2013, pp. 732--736.

\bibitem{data2}
Art-sci center, \uppercase{U}niversity of \uppercase{C}alifornia,
  \url{http://artsci.ucla.edu/birds/database.html/}, accessed: 2016-07-10.

\bibitem{data1}
Cassin's vireo recordings,
  \url{http://taylor0.biology.ucla.edu/al/bioacoustics/}, accessed: 2016-03-20.

\bibitem{fant}
Filtering and noise adding tool, \url{http://dnt.kr.hs-niederrhein.de/},
  accessed: 2016-11-14.

\bibitem{freesound}
Freesound, \url{http://freesound.org/}, accessed: 2017-3-13.

\bibitem{xu_attention}
Y.~Xu, Q.~Kong, Q.~Huang, W.~Wang, M.~D. Plumbley, Attention and localization
  based on a deep convolutional recurrent model for weakly supervised audio
  tagging, in: Interspeech, 2017, pp. 1267--1271.

\bibitem{lee2001algorithms}
D.~D. Lee, H.~S. Seung, Algorithms for non-negative matrix factorization, in:
  Advances in neural information processing systems, 2001, pp. 556--562.

\bibitem{deep_learning_book}
I.~Goodfellow, Y.~Bengio, A.~Courville, Deep Learning, MIT Press, 2016,
  \url{http://www.deeplearningbook.org}.

\end{thebibliography}
\end{document}